\documentclass[a4paper,10pt,aps,pre,twocolumn,amsmath,amssymb,longbibliography]{revtex4-1}

\usepackage{graphicx,color}
\usepackage[utf8]{inputenc}
\usepackage[colorlinks=true,citecolor=blue]{hyperref}

\newcommand{\ahum}[1]{``#1''}
\newcommand{\eq}[1]{Eq.~(\ref{#1})}
\newcommand{\fig}[1]{Fig.~\ref{#1}}

\newcommand{\olcite}[1]{Ref.~\cite{#1}}
\newcommand{\avg}[1]{\langle #1 \rangle}
\newcommand{\etal}{{\it et al.}}

\newcommand{\tauS}{\tau_{\rm slip}}
\newcommand{\Tsub}{T_{\rm sub}}
\newcommand{\Tlan}{T_{\rm lan}}
\newcommand{\fric}{F_{\rm fr}}

\begin{document}

\title{On the connection between sliding friction and phonon lifetimes: Thermostat induced 
thermolubricity effects in molecular dynamics simulations}

\author{Richard L. C. Vink}

\affiliation{Institute of Materials Physics, University of Goettingen, 
Friedrich-Hund-Platz~1, D-37077 Goettingen, Germany}

\date{\today}

\begin{abstract} A typical nanotribology simulation setup is the semi-infinite substrate, featuring 
a sliding bead on top, and with the lower substrate layers thermostatted to control temperature. A 
challenge is dealing with phonons that backreflect from the substrate lower boundary, as these will 
artificially reduce the friction $\fric$ acting on the sliding bead. One proposed solution is to use 
a Langevin thermostat, operating at temperature $\Tlan$, and with the corresponding damping 
parameter, $\gamma$, optimally tuned such that $\fric$ is maximized [Benassi \etal, Phys.~Rev.~B 
{\bf 82}, 081401 (2010)]. In this paper, the method is revisited, and related to the substrate 
phonon lifetime, the substrate temperature $\Tsub$, and the sliding speed. At low sliding speed, 
where the time between stick-slip events is large compared to the phonon lifetime, we do not observe 
much dependence of $\fric$ on $\gamma$, and here thermostat tuning is not required. At high sliding 
speed, upon varying $\gamma$, we confirm the aforementioned friction maximum, but also observe a 
pronounced minimum in $\Tsub$, which here deviates from $\Tlan$. For substrate particle interactions 
that are strongly anharmonic, the variation of $\fric$ with $\gamma$ can be understood as a 
manifestation of thermolubricity, backreflections being essentially unimportant. In contrast, for 
harmonic interactions, where phonon lifetimes become very long, $\fric$ is strongly affected by 
backreflecting phonons, though not enough to overturn thermolubricity. \end{abstract}

\maketitle

\section{Introduction}

\begin{figure}
\begin{center}
\includegraphics[width=0.9\columnwidth]{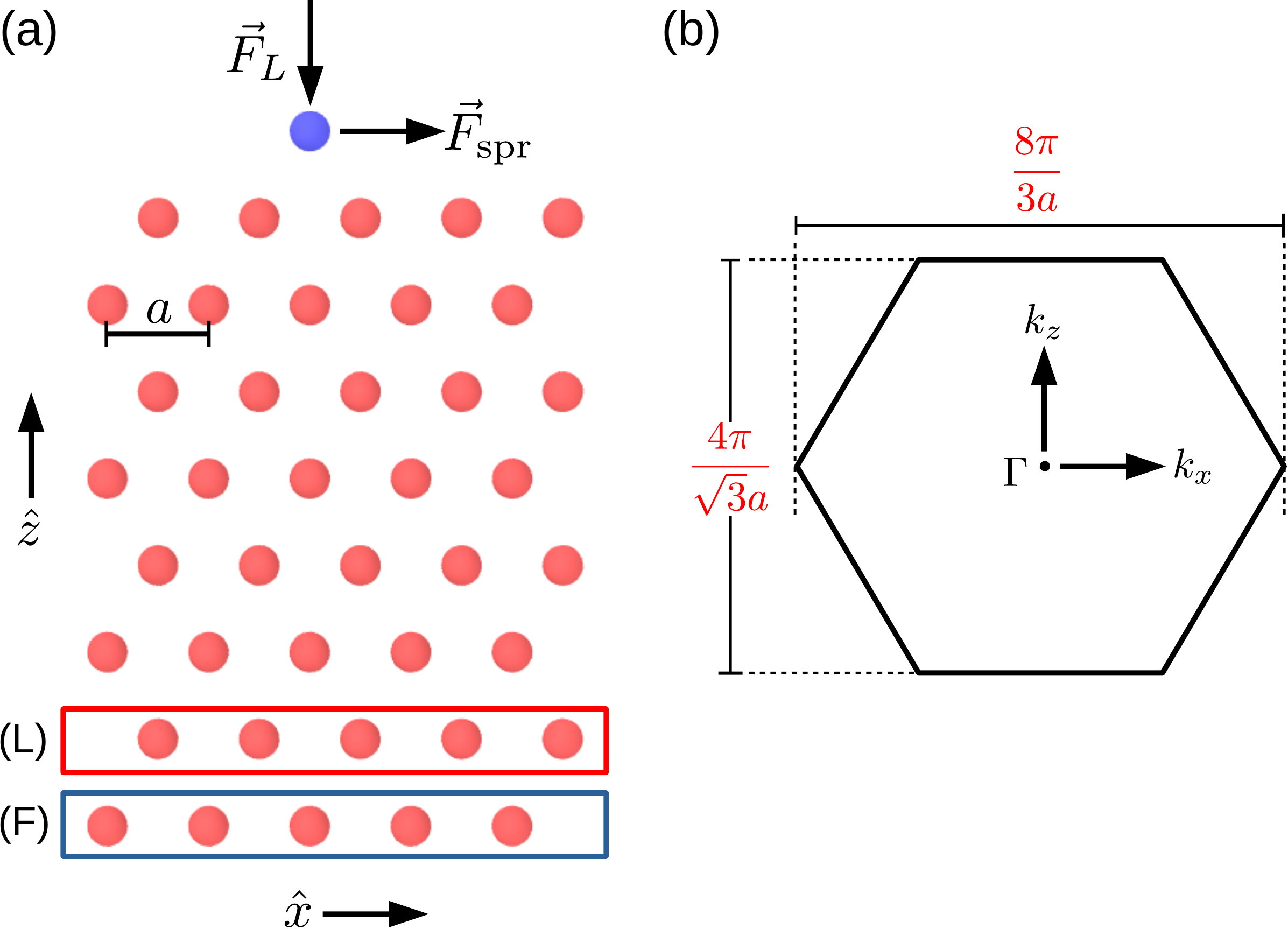}

\caption{\label{fig1} (a) Cartoon of a basic nanotribology MD simulation setup, also used in this 
work. An AFM bead (top particle), to which a vertical load $\vec{F}_L$ is applied, is dragged over a 
substrate (nearest neighbor distance $a$) into the lateral direction $\hat{x}$ by a force 
$\vec{F}_{\rm spr}$. To prevent the entire substrate from sliding, the particles in the lowest layer 
(F) are kept fixed during the simulation. To control the temperature, a (Langevin) thermostat is 
applied to the second layer (L). (b) Sketch of the first Brillouin zone (FBZ) of the hexagonal 
lattice used in the simulations. The FBZ is a regular hexagon, with size as indicated. The center of 
the zone marks the $\Gamma$ point.}

\end{center}
\end{figure}

Molecular Dynamics (MD) simulations are an established tool in the field of (nano)tribology. 
Relatively straightforward to setup, these simulations can be used to study the dynamics of a 
contact of some kind (e.g.~the tip of an atomic force microscope) sliding across a surface. A 
practical problem facing all these simulations is how to remove the excess energy pumped into the 
system at the sliding contact, for, if this energy is not removed, the system temperature would 
steadily rise. The problem has received considerable attention~\cite{10.1103/physrevb.82.081401, 
10.1088/0953-8984/22/7/074205, 10.1007/s11249-012-9936-5, 10.1103/physrevb.78.094304, 
10.1103/physrevb.76.104107, 10.1002/anie.201400066} and one practical solution that has emerged is 
to remove the excess energy via a thermostat acting on the lower part of the substrate, as depicted 
schematically in \fig{fig1}(a). The thermostat will, of course, alter the true dynamics of the 
system in the lower region, but if the simulation cell is large enough, one might hope that the 
dynamics in the upper region remains unaffected, facilitating meaningful friction measurements.

Alas, simulations using the geometry of \fig{fig1}(a) unambiguously show that the measured friction 
force $\fric$ depends quite sensitively on thermostat details~\cite{10.1103/physrevb.82.081401, 
10.1007/s11249-012-9936-5, 10.3762/bjnano.8.218}. In these works, this dependence is mainly 
attributed to acoustic phonons that get \ahum{backreflected} from the lower simulation box boundary. 
In very simple terms, as the bead is dragged across the surface, acoustic phonons are generated in 
the underlying substrate. The fixed atoms at the bottom of the simulation cell act as a 
\ahum{mirror} reflecting these phonons back toward the bead, potentially allowing the bead to regain 
some of its energy, implying that the measured friction force will be \ahum{too small}, compared to 
what it would be in the infinite system without backreflections. The key point to note next is that, 
by tuning the thermostat, which acts on the particles directly above the frozen bottom layer, the 
degree of phonon backreflection can be regulated to some 
extent~\cite{10.1088/0953-8984/22/7/074205}. Hence, it seems logical to tune the thermostat such 
that the measured friction force is maximized, since, under this condition, the artificial reduction 
of friction due to phonon backreflections must necessarily be minimized. Indeed, taking for 
concreteness a Langevin thermostat, where the adjustable parameter is the damping coefficient 
$\gamma$, the existence of an \ahum{optimal} damping coefficient, where the friction force reaches a 
maximum, was strikingly confirmed~\cite{10.1103/physrevb.82.081401, 10.1007/s11249-012-9936-5, 
10.3762/bjnano.8.218}.

The purpose of this paper is to relate these findings to the phonon properties of the underlying 
substrate, specializing to the regime where the motion of the bead is of the stick-slip type. Quite 
surprisingly, we confirm the previous findings~\cite{10.1103/physrevb.82.081401, 
10.1007/s11249-012-9936-5, 10.3762/bjnano.8.218} in certain dynamic regimes, but not in others. To 
be precise: the friction maximum upon variation of the thermostat damping parameter $\gamma$ is only 
observed at high sliding speeds; at low sliding speeds, the friction force does not reveal any 
systematic $\gamma$ dependence. The defining criterion turns out to be the typical phonon lifetime 
in the substrate: When the latter is small compared to the time between single stick-slip events, 
the influence of the thermostat on friction vanishes. In the opposite limit, friction depends quite 
sensitively on thermostat details, but the extent to which this impedes meaningful measurements 
depends sensitively on the degree of anharmonicity in the substrate particle interactions. For 
anharmonic substrates, the dependence of friction on thermostat details is, for the most part, 
explained by thermolubricity~\cite{10.1103/physrevlett.91.084502, 10.1103/physrevlett.104.256101}.

We will, in what follows, present the results of MD simulations leading to these conclusions. Our 
simulations are based on a two-dimensional (2D) model system, described in Section~II, which is 
similar in spirit to that of \olcite{10.1103/physrevb.82.081401}. In Section~III, we present the 
corresponding friction measurements, as well as the substrate phonon properties required to explain 
these measurements. The relation between friction and phonon properties, in particular the phonon 
lifetime, is discussed in Section~IV. We end with a summary and some recommendations in Section~V.

\section{Model, Methods, and Definitions}

We simulate the 2D setup of \fig{fig1}(a). The substrate is a hexagonal lattice, whose primitive 
cell is spanned by the vectors $\vec{a}_1 = a \hat{x}$ and $\vec{a}_2 = \frac{a}{2}(\hat{x} + 
\sqrt{3}\hat{z})$, each cell containing exactly one particle of mass $m$, with $a$ the lattice 
constant. The cell is replicated $\sqrt{N}$ times in both directions $\vec{a}_{1,2}$, $N=1600$ being 
the particle number (the simulation box is thus triclinic). We apply periodic boundary 
conditions in the (horizontal) $x$-direction, but not in the (vertical) $z$-direction. Between 
nearest neighboring particles, springs are attached, the energy of a single spring being $u_{\rm 
spr} (r) = \sum_{n=2}^4 \alpha_n (r-a)^n$, where $r$ is the spring length. To anchor the substrate, 
the lowest row of particles have their positions fixed.

A Langevin thermostat~\cite{10.1103/physrevb.17.1302, 10.1142/s0129183191001037} is applied to the 
row of particles directly above the fixed layer. To the forces acting on these particles, damping 
and stochastic terms are added:
\begin{equation} 
\label{eq:lan}
 \vec{F}_{\rm lan} = -\gamma m \, \vec{v} 
 + \sqrt{(24 k_B \Tlan \gamma m/ \delta t)} \, \vec{h} \quad,
\end{equation}
in addition to the forces arising from the springs. Here, $\gamma$ is the damping parameter, 
$\delta t$ the MD integration time step, $k_B$ the Boltzmann constant, $\vec{v}$ the particle 
velocity, and $\vec{h}$ a two-dimensional vector with components drawn uniformly from $[-0.5:0.5]$. 
For each thermostatted particle, at each MD step, a new vector $\vec{h}$ is to be generated! 
Note that the Langevin limits $\gamma \to 0$ and $\gamma \to \infty$ are equivalent, both 
corresponding to a system without thermostat. For $\gamma=0$ this is intuitively clear, since here 
$\vec{F}_{\rm lan}=0$. For $\gamma \to \infty$, the stochastic term, $\propto \sqrt{\gamma}$, 
becomes negligible compared to the viscous term, $\propto \gamma$, leading to an infinite viscous 
force, which will impede any motion of the thermostat particles. Hence, $\gamma \to \infty$ is 
similar to $\gamma=0$, but with an extra layer of frozen particles (direct simulations in the regime 
$\gamma \to \infty$ are, however, numerically challenging, since a progressively smaller MD timestep 
$\delta t$ is then required).

The temperature $\Tlan$ appearing in \eq{eq:lan} is the thermostat temperature. In thermal 
equilibrium, this is also the temperature the substrate will adopt, $\Tsub=\Tlan$, with $\Tsub$ 
computed directly from the particle kinetic energy, $k_B T_{\rm sub} = (2/d) \avg{K}$, with $d$ the 
spatial dimension, and where $\avg{K}$ is the average kinetic energy of the substrate particles 
(when computing $\avg{K}$, we exclude the bead particle, as well as the frozen and thermostatted 
substrate layers). We already announce here that, in non-equilibrium situations, $\Tsub$ and $\Tlan$ 
can be very different!

On top of the substrate, a bead particle is placed, having the same mass $m$ as the substrate 
particles. The bead is subjected to a vertical load, $\vec{F}_L = -L \hat{z}$, pressing it down onto 
the underlying substrate. In addition, the bead is attached to one end of a harmonic spring (spring 
constant $k$, zero rest length), while the other end of the spring is dragged to the right with 
constant velocity~$v$. The spring is assumed to act only in the lateral $x$-direction, that is, it 
induces a force $\vec{F}_{\rm spr} = k[X(t) - v t] \hat{x}$ onto the bead, where $t$ is the time, 
and $X(t)$ the $x$-coordinate of the AFM bead at time $t$. The average friction force magnitude is 
then obtained by time averaging the lateral spring force: $F_{\rm fr} = k \avg{ X(t) - v t }_t$.

The interaction energy between the AFM bead and the underlying substrate particles is taken to be a 
sum of (short-ranged) pair potentials, $U_{\rm afm} = \sum_s^\prime u_{\rm afm} (R_s)$, with the sum 
over all substrate particles $s$ whose distance $R_s$ from the AFM bead is smaller than a specified 
cutoff distance: $R_s < R_c=2.5a$. The pair potential is of the $(12,6)$ Lennard-Jones (LJ) form, 
with added linear term:
\begin{equation}
\label{eq1}
 u_{\rm afm} (r) = c_1 \epsilon \left[ \left(\frac{c_2}{r}\right)^{12} -
\left(\frac{c_2}{r}\right)^6  \right] + c_3 r + c_4 \quad,
\end{equation}
where $\epsilon$ sets the energy scale, and with the constants $c_i$ chosen such that the minimum of 
the pair potential is located at $r=a$, with corresponding value $u_{\rm afm}(a)=-0.6\epsilon$, 
while at the cutoff $u_{\rm afm}( R_c ) = u_{\rm afm}'( R_c ) = 0$, i.e.~there is no force 
discontinuity.

We adopt LJ units throughout: $a = \epsilon = m = k_B \equiv 1$. The vertical load is set to $L=10$, 
the AFM spring constant to $k=5$. The time evolution of the system is obtained via standard 
molecular dynamics, using the velocity-verlet algorithm as implemented in 
LAMMPS~\cite{10.1006/jcph.1995.1039}, with integration time step~$\delta t=0.001$. The sliding 
velocity $v$ will be varied, but we take care to remain in the \ahum{stick-slip} regime, the AFM 
bead thus spending most of its time in low-energy positions on the substrate, the transitions 
between such positions being rapid. Since our primary interest is friction on solid supports, we 
consider low temperatures only, typically choosing $\Tlan=0.035$, which is well below 
melting~\cite{10.1103/physrevb.83.214108}.

In what follows, we shall refer to harmonic and anharmonic substrates. For the harmonic substrate, 
the spring energy parameters $\alpha_2=36,\alpha_3=\alpha_4=0$; for the anharmonic substrate 
$\alpha_2=36, \alpha_3=-252, \alpha_4=1113$. These parameters correspond to a Taylor expansion of a 
$(12,6)$ LJ potential around its minimum, with well-depth $\epsilon=1$, and minimum 
located at $r=1$. We emphasize that both substrates are strictly speaking anharmonic, since, even 
for the harmonic version, the potential energy is not quadratic in the particle displacements. 
However, owing to the explicit absence of 3rd and 4th order terms, anharmonicity effects should be 
much weaker, which suffices for our purposes.

\subsection{Phonon properties}

In our analysis, we will relate our friction measurements to the vibrational properties of the 
substrate using the language of phonons. Each phonon is characterized by a wavevector 
$\vec{k}=(k_x,k_z)$ and polarization $p$. In a finite system, the number of wavevectors inside the 
first Brillouin zone (FBZ) equals the number of primitive unit cells, which here equals the number 
of particles $N$, since the hexagonal lattice has a 1-atom basis. In \fig{fig1}(b), we show the FBZ 
of the hexagonal lattice, properly oriented, i.e.~corresponding to the primitive vectors 
$\vec{a}_{1,2}$ of the lattice in real space. The center of the FBZ marks the $\Gamma$ point: 
$\vec{k}=(0,0)$. Each wavevector yields two polarizations, longitudinal-acoustic ($p=\rm LA$) and 
transverse-acoustic ($p=\rm TA$), the respective speeds of sound being $c_{\rm TA} = (\sqrt{3}/2) a 
\sqrt{\alpha_2/m}$ and $c_{\rm LA} = (3/2) a \sqrt{\alpha_2/m}$.

For each phonon propagating in the direction $\vec{k}$, we introduce its lifetime $\tau(\vec{k},p)$, 
and energy $E(\vec{k},p)$, with $p \in \rm TA,LA$. These quantities can be obtained from the 
substrate particle positions and velocities in the MD trajectory (details in Appendix). Phonons with 
short lifetimes $\tau(\vec{k},p)$ are dissipating, since these phonons quickly distribute their 
energy over other phonon modes, leading to rapid thermalization of the entire phonon population. 
Analogously, phonons with long lifetimes $\tau(\vec{k},p)$ are non-dissipating, since these can 
store their energy over longer times, delaying thermalization. In applications, thermalization of 
the substrate, i.e.~the conversion of the bead's kinetic energy into heat, is typically undesirable, 
and so to understand how the sliding motion of the bead couples to the phonons of the underlying 
substrate, is very important.

\section{Results}

\subsection{Friction measurements}

\begin{figure}
\begin{center}
\includegraphics[width=0.85\columnwidth]{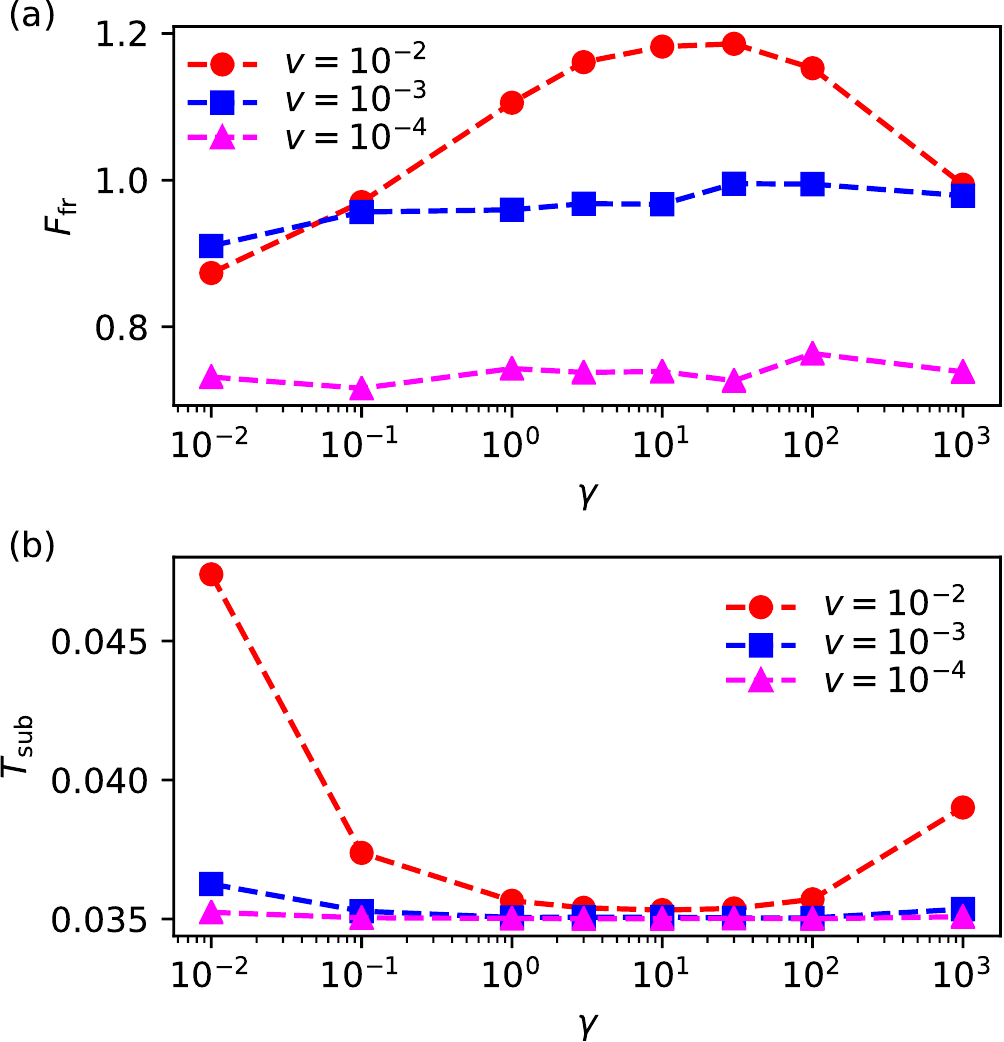}
\caption{\label{fig2} Variation of (a) the average friction force $\fric$, and (b) substrate 
temperature $\Tsub$, with the Langevin parameter $\gamma$ for the harmonic substrate, for various 
sliding speeds $v$, as indicated. The used thermostat temperature $\Tlan=0.035$.}
\end{center}
\end{figure}

\begin{figure}
\begin{center}
\includegraphics[width=0.85\columnwidth]{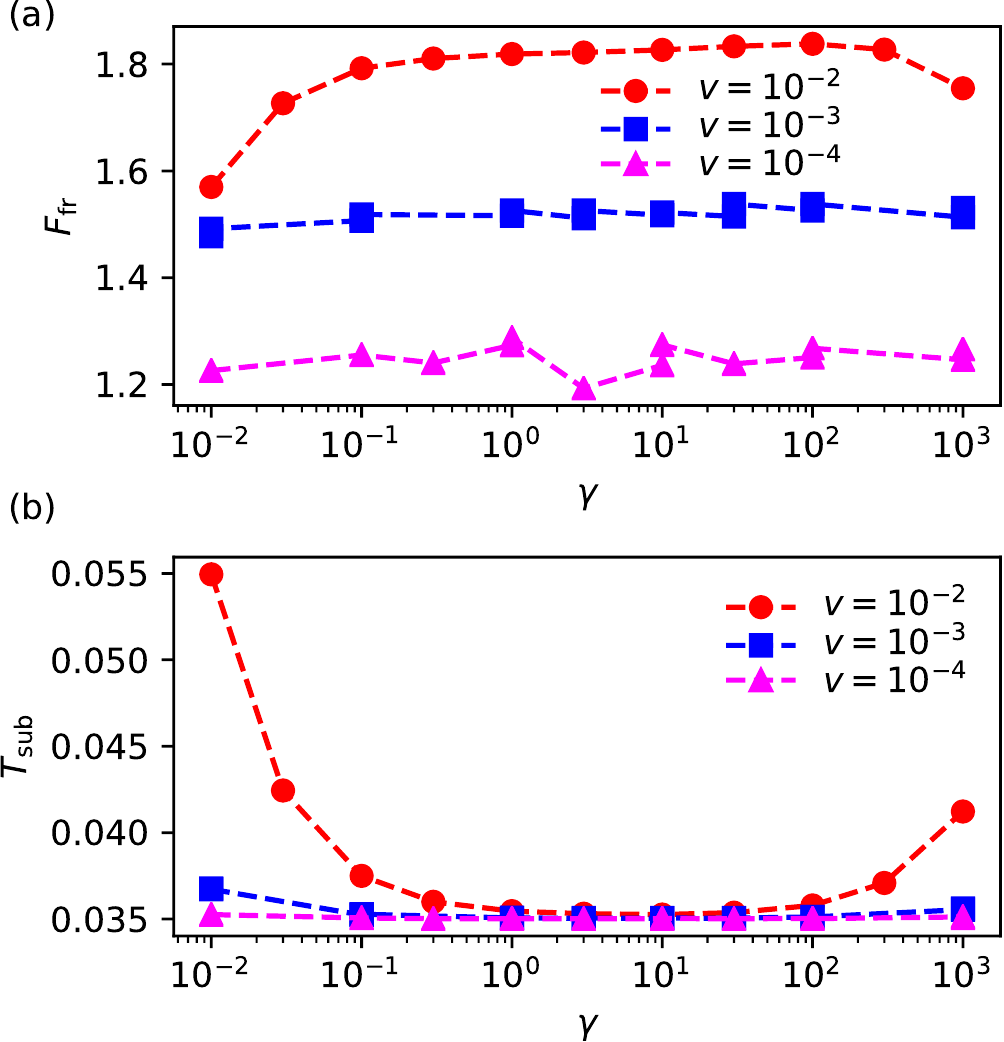}
\caption{\label{fig3} The analogue of \fig{fig2} for the anharmonic substrate.}
\end{center}
\end{figure}

We first measure how the friction force, $\fric$, depends on the Langevin thermostat parameter, 
$\gamma$, as well as on the sliding speed, $v$. The temperature of the Langevin thermostat 
$\Tlan=0.035$. However, this being a non-equilibrium situation, the actual temperature of the 
substrate, $\Tsub$, may well deviate from $\Tlan$. For each friction measurement, $\sim 20 \cdot 
10^6$ MD steps were applied to bring the system into a steady state, followed by production runs of 
at least $500 \cdot 10^6$ MD steps, during which the average friction force was measured. In 
\fig{fig2}(a), we show, for the harmonic substrate, the dependence of $\fric$ on $\gamma$ for three 
values of the sliding speed, $v$, as indicated. The lower panel, \fig{fig2}(b), shows the 
variation of $\Tsub$ with $\gamma$. For the anharmonic substrate, the analogous analysis is 
presented in \fig{fig3}.

At first sight, the data of \fig{fig2} and \fig{fig3} look very similar. For both the harmonic and 
anharmonic substrate, at the highest sliding speed $v=0.01$ considered (which remains well below 
the speed of sound), the curve of $\fric$ versus $\gamma$ reveals a maximum. The existence of the 
friction maximum agrees with previous studies~\cite{10.1103/physrevb.82.081401, 
10.1007/s11249-012-9936-5, 10.3762/bjnano.8.218}, and is attributed to phonon backreflections, which 
are maximally suppressed at the friction maximum. However, still considering the highest sliding 
speed, the data also reveal that the substrate temperature, $\Tsub$, varies with $\gamma$, reaching, 
in fact, a minimum at the friction maximum (with the value of $\Tsub$ then being close to $\Tlan$ of 
the thermostat). Since friction is known to depend on temperature as well, an effect called 
thermolubricity~\cite{10.1103/physrevlett.91.084502, 10.1103/physrevlett.104.256101}, our data 
indicate that the observed friction maximum could be the manifestation of two effects, namely, 
suppression of phonon backreflections {\it and} reducing the substrate temperature, both of which 
would increase friction.

Interestingly, lowering the sliding speed $v$, these effects gradually vanish, $\fric$ and $\Tsub$ 
then being essentially independent of $\gamma$, with $\Tsub$ close to $\Tlan$ of the thermostat. For 
both the harmonic and anharmonic substrate, reducing the sliding speed leads to lower friction. Note 
also that, for the anharmonic substrate, friction significantly exceeds that of the harmonic 
substrate.

\begin{figure*}
\begin{center}
\includegraphics[width=2\columnwidth]{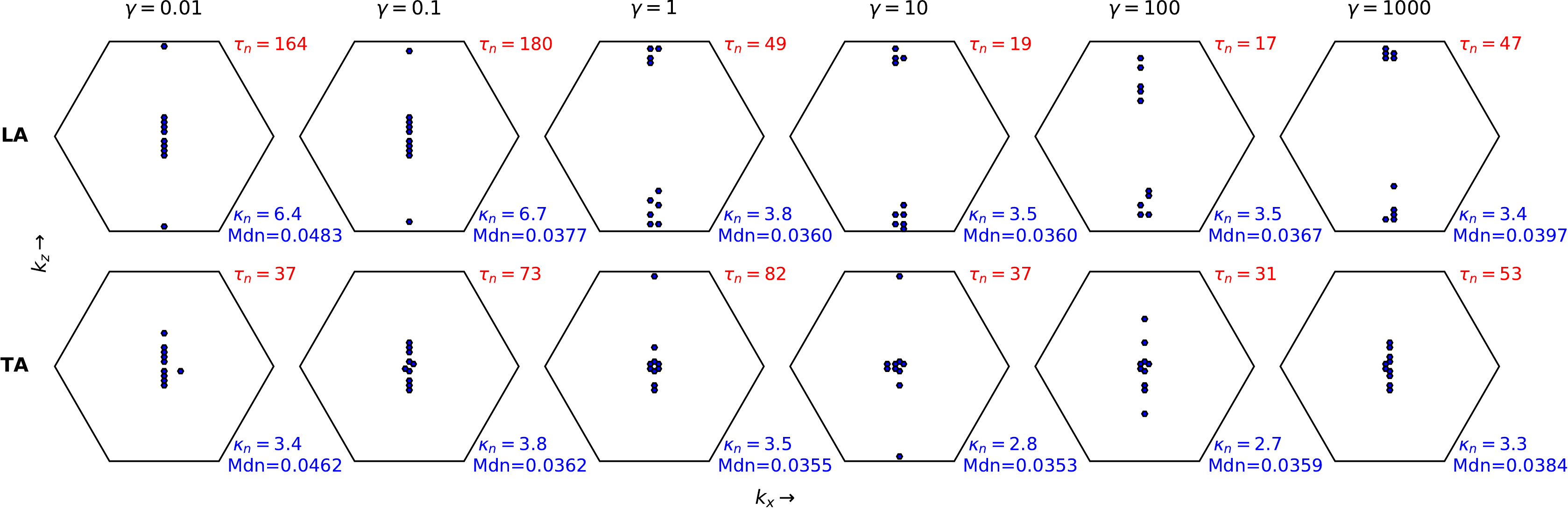}

\caption{\label{fig4} Graphical representation showing the set $S_n$ of $n=10$ maximally excited 
phonons, for the harmonic substrate, for various values of the Langevin thermostat parameter 
$\gamma$ (columns). Each hexagon represents the FBZ, the symbols inside mark the wavevectors 
$\vec{k}=(k_x,k_z)$ of the modes in $S_n$, for LA polarization (top row), and TA polarization 
(bottom row). Also indicated for each measurement is the average phonon lifetime $\tau_n$, and the 
average phonon excitation $\kappa_n$, with the average taken over the set $S_n$, see details in 
text. In addition, we indicate the median ${\rm Mdn}[E(\vec{k},p)]$ of the entire phonon population 
(Mdn). All data use thermostat temperature $\Tlan=0.035$, sliding speed $v=0.01$.}

\end{center}
\end{figure*}

\begin{figure*}
\begin{center}
\includegraphics[width=2\columnwidth]{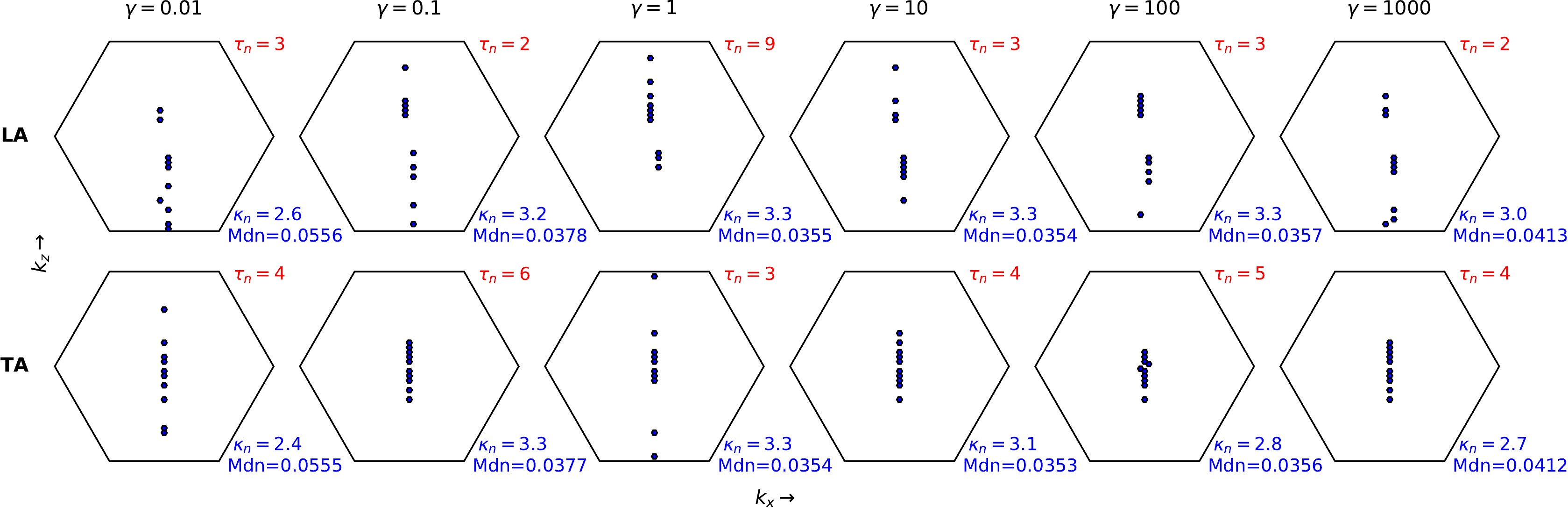}
\caption{\label{fig5} The analogue of \fig{fig4} for the anharmonic substrate.}
\end{center}
\end{figure*}

\subsection{Effect of sliding on substrate phonons}

We now identify the phonon modes most excited by the sliding bead, at the highest considered sliding 
speed $v=0.01$, thermostat temperature $\Tlan=0.035$, for various values of $\gamma$, and for both 
substrate types (harmonic, anharmonic). To this end, we measure the time-averaged energy 
$E(\vec{k},p)$ of the phonon characterized by the wavevector $\vec{k}$ and polarization $p \in \rm 
TA,LA$. As measure for the degree of excitation, we use the quantity $\kappa(\vec{k},p) \equiv 
E(\vec{k},p) / {\rm Mdn}[E(\vec{k},p)]$, where ${\rm Mdn}[E(\vec{k},p)]$ is the {\it median} (not 
mean) of the obtained $E(\vec{k},p)$ values. In thermal equilibrium, at low temperature, one may 
assume equipartition approximately holds, in which case all phonon modes should have the same 
energy: $E(\vec{k},p) = {\rm Mdn}[E(\vec{k},p)] = k_B \Tsub$, implying $\kappa(\vec{k},p)=1$. Under 
driving, we still expect a substantial fraction of the phonon population to be thermalized, with the 
exception of modes that couple strongly to the sliding motion of the bead, whose energy should then 
exceed the thermal value, implying $\kappa(\vec{k},p)>1$. To confirm these ideas, we identify, for 
each polarization separately, the set $S_n$ of $n=10$ phonons with the largest value of 
$\kappa(\vec{k},p)$, and mark the corresponding wavevectors in the FBZ. For these phonons, we 
additionally compute the average lifetime, $\tau_n = (1/n) \sum^\prime \tau (\vec{k},p)$, and the 
average excitation, $\kappa_n = (1/n) \sum^\prime \kappa(\vec{k},p)$, where the sum is over all 
phonons in the set $S_n$. For the harmonic (anharmonic) substrate, the result of this procedure is 
shown in \fig{fig4} [\fig{fig5}].

For the harmonic substrate, $\tau_n$ varies strongly with $\gamma$ [\fig{fig4}]. For both 
polarizations, by increasing $\gamma$, $\tau_n$ first reaches a maximum, then decreases to a minimum 
at $\gamma \sim 10-100$, which roughly corresponds to the friction maximum [\fig{fig2}(a)]. In 
addition, for small $\gamma \leq 0.1$, we observe a vertical band around the $\Gamma$ point of LA 
phonons with very long lifetimes; by increasing $\gamma$, this band disappears. Apparently, the 
effect of increasing $\gamma$ is to drastically reduce the lifetime of long wavelength, vertically 
propagating, LA phonons. Also of interest is the variation of ${\rm Mdn}[E(\vec{k},p)]$ with 
$\gamma$. As can be seen by comparing to \fig{fig2}(b), the median rather closely follows the 
substrate temperature, $k_B \Tsub$, confirming that many phonons are still thermalized. However, 
phonons inside the sets $S_n$, i.e.~those which couple most strongly to the bead, have energies that 
exceed the thermal value by factors typically $\kappa_n \sim 3$, the exception being LA phonons 
at small $\gamma$, for which the enhancement is significantly larger, $\kappa_n > 6$.

For the anharmonic substrate, we observe a much weaker dependence of $\tau_n$ on $\gamma$, whose 
value, in comparison to the harmonic substrate, is now much smaller [\fig{fig5}]. In contrast to the 
harmonic substrate, vertical bands in the FBZ persist for all values of $\gamma$, i.e.~a preferred 
suppression of the lifetime of vertically propagating LA phonons upon increasing $\gamma$ does not 
take place. The typical phonon enhancement $\kappa_n \sim 3$ for all values of $\gamma$. As before, 
the median closely follows the substrate temperature, $k_B \Tsub$, see \fig{fig3}(b). There is, 
however, one subtle difference: For the harmonic substrate, ${\rm Mdn}[E(\vec{k},p)]$ for TA and LA 
modes differs by about $\pm 0.001$, while for the anharmonic substrate, the difference is only $\pm 
0.0001$, i.e.~ten times smaller. This indicates that the deviation from thermal equilibrium is 
largest for the harmonic substrate.

\begin{figure}
\begin{center}
\includegraphics[width=\columnwidth]{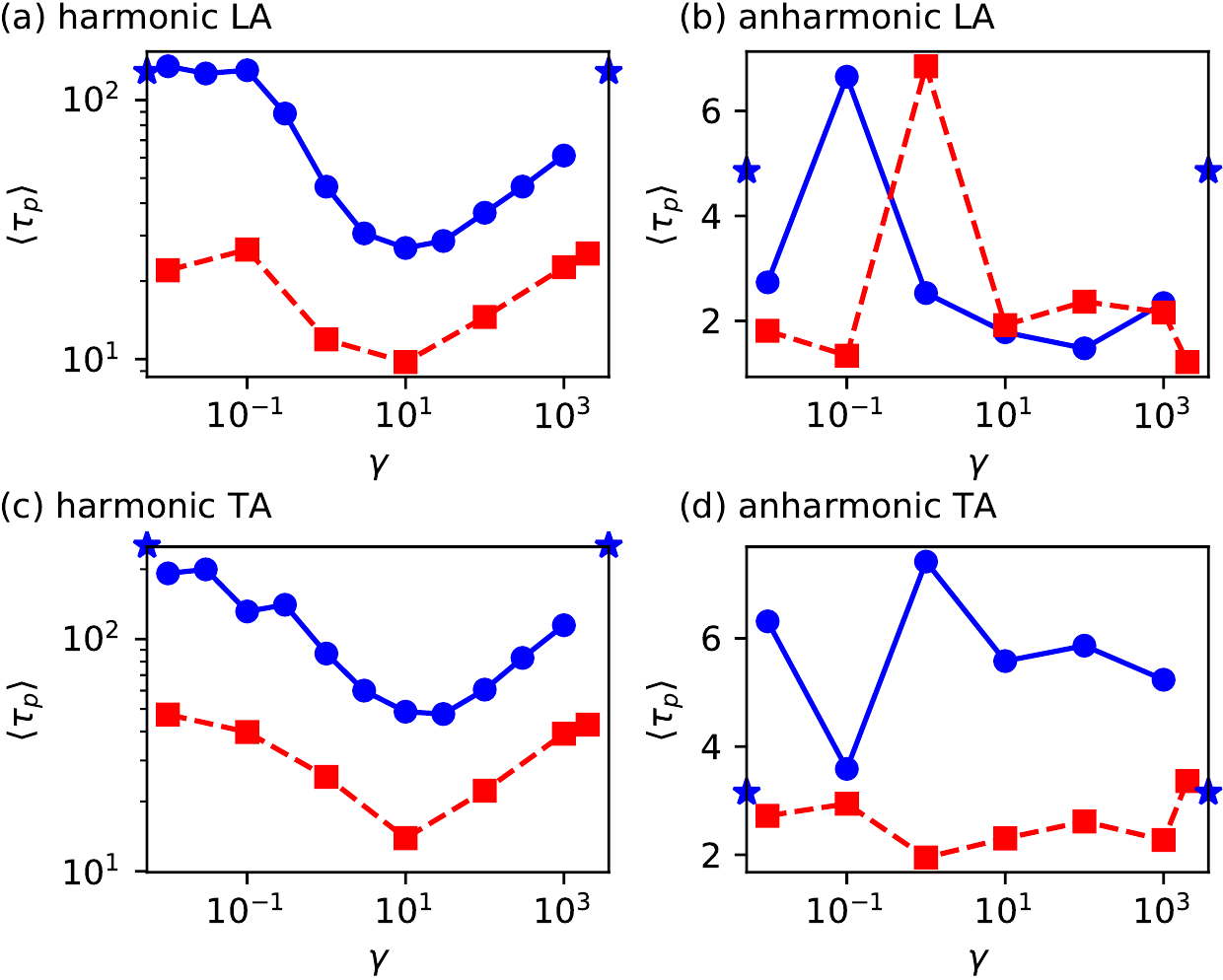}

\caption{\label{fig6} Variation of the typical phonon lifetime $\avg{\tau_p}$ with $\gamma$ obtained 
in equilibrium (circles) and under sliding (squares). The star symbols show the lifetime obtained in 
equilibrium without Langevin thermostat. Results are sorted by substrate type and polarization, see 
the plot titles. Note that, for the harmonic substrate, the vertical scale is logarithmic!}

\end{center}
\end{figure}

\subsection{Typical phonon lifetimes}

Finally, we still measure the typical substrate phonon lifetime, $\avg{\tau_p}$, where typical means 
an average over all wavevectors (details in Appendix). When computing $\avg{\tau_p}$, we disregard 
whether the phonon modes being averaged over are actually excited by the bead, which is the main 
difference from $\tau_n$ defined previously. In \fig{fig6}, we show $\avg{\tau_p}$, for both 
substrate types, and for both polarizations $p$. Results are shown for the equilibrium case ($v=0$: 
circles), and under sliding ($v=0.01$: squares) using again $\Tlan = 0.035$. For the equilibrium 
case, after initial equilibration, active thermostatting is not required to maintain the 
temperature. In this case, we can \ahum{turn off} the Langevin thermostat, yielding the star symbols 
in \fig{fig6}.

For the harmonic substrate, \fig{fig6} reveals a strong dependence of $\avg{\tau_p}$ on $\gamma$, 
reaching a minimum at $\gamma \sim 10$, which again is close to the friction maximum 
[cf.~\fig{fig2}(a)]. In addition, upon sliding, $\avg{\tau_p}$ is significantly reduced from its 
equilibrium value. Apparently, the sliding bead introduces extra \ahum{noise} into the substrate, in 
addition to that of the Langevin stochastic term, promoting phonon mixing. We also observe that, in 
the limits $\gamma \to 0$ and $\gamma \to \infty$, the equilibrium values (circles) approach the 
star symbols obtained without Langevin thermostat.

For the anharmonic substrate, compared to the harmonic one, \fig{fig6} reveals an overall much 
smaller phonon lifetime: Only for $\gamma \sim 10$ and under sliding, do the values become somewhat 
close. Furthermore, there are pronounced qualitative differences. For LA modes, neither $\gamma$ nor 
sliding systematically affect $\avg{\tau_p}$, the corresponding values remaining close (in absolute 
terms) to those obtained without Langevin thermostat (stars). For TA modes, a systematic dependence 
on $\gamma$ remains absent, but sliding does appear to slightly reduce $\avg{\tau_p}$. Still, in 
absolute terms, also for TA modes, the observed lifetimes are all rather similar, remaining close to 
those obtained without Langevin thermostat.

\section{Discussion}

\subsection{Friction at low sliding speeds}

At the lowest considered sliding speed, $v=10^{-4}$, for both substrate types, neither $\fric$ nor 
$\Tsub$ show any appreciable dependence on the Langevin damping parameter $\gamma$ [\fig{fig2} and 
\fig{fig3}]. This can be understood from the typical phonon lifetime $\avg{\tau_p}$. At $v=10^{-4}$, 
the time between stick-slip events $\tauS = a/v = 10000$~LJ time units, which far exceeds 
$\avg{\tau_p}$, for all values of $\gamma$, and for both substrate types [\fig{fig6}]. The reduction 
of friction due to phonon backreflections thus cannot occur since any coherence between the bead and 
the substrate phonons generated during slip, will long have decayed by the time of the next slip 
event. Consecutive slip events are thus uncorrelated, and, at the start of each such event, the 
underlying substrate in a state of thermal equilibrium. Friction simulations in this regime are thus 
relatively straightforward, as essentially any value of $\gamma$ suffices, always yielding a 
friction value corresponding to the temperature of the thermostat, which then equals that of the 
substrate: $\Tlan = \Tsub$.

At the intermediate sliding speed, $v=10^{-3}$, $\tauS=1000$~LJ time units. For the anharmonic 
substrate, this still far exceeds $\avg{\tau_p}$. In line with our argumentation, the corresponding 
friction data do not show any appreciable dependence on $\gamma$ [\fig{fig3}(a)]. For the harmonic 
substrate, at small $\gamma$, $\avg{\tau_p}$ is still below $\tauS$, but not that much lower, which 
means there could be some influence of backreflections. Consistent with this interpretation, a 
slight decrease of $\fric$ at small $\gamma$ is visible then [\fig{fig2}(a)].

\subsection{Friction at high sliding speed}

\begin{figure}
\begin{center}
\includegraphics[width=0.85\columnwidth]{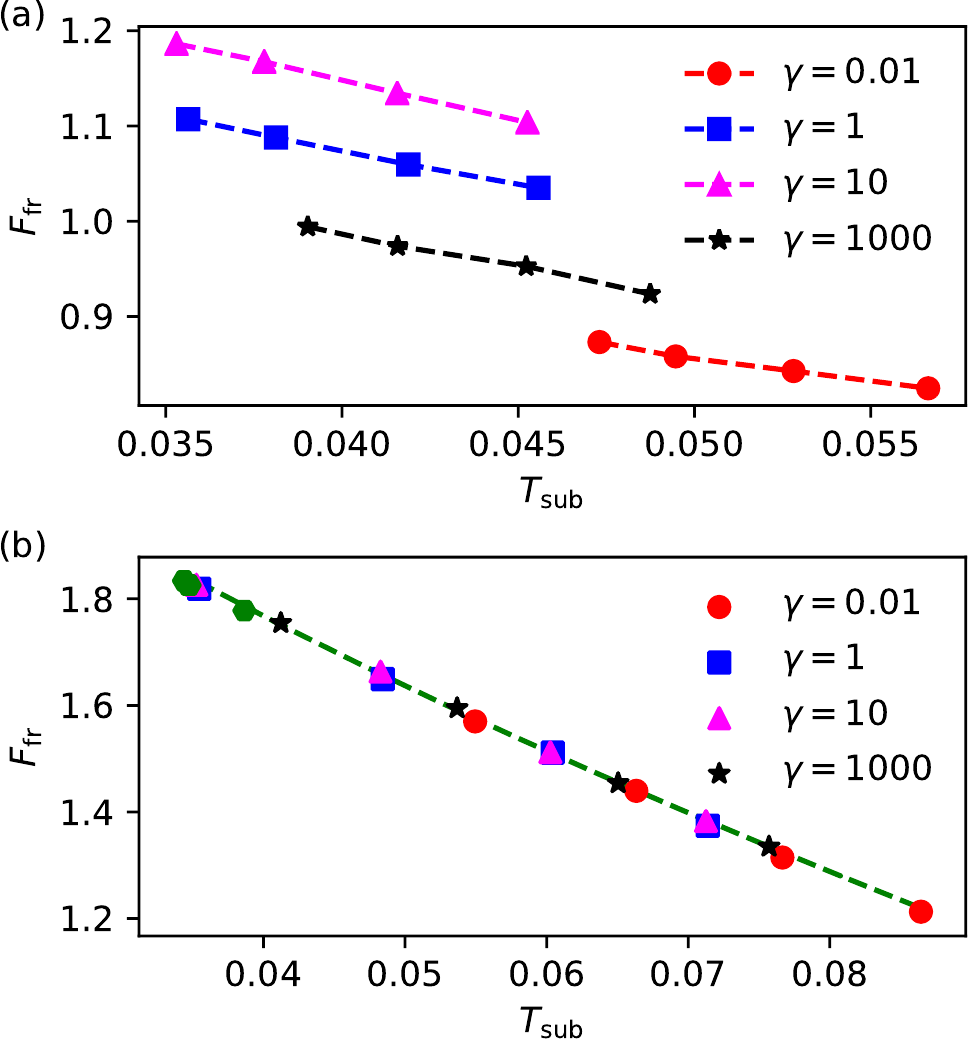}

\caption{\label{fig7} Friction, $\fric$, versus substrate temperature, $\Tsub$, for (a) harmonic and 
(b) anharmonic substrates, for different values of the Langevin parameter, $\gamma$, as indicated 
(for the harmonic substrate, data correspond to $\Tlan=0.035, 0.0375, 0.0413, 0.045$; for the 
anharmonic substrate, $\Tlan=0.035, 0.048, 0.06, 0.071$). In addition, for the anharmonic substrate, 
results obtained using a velocity rescaling thermostat are included (hexagon symbols).}

\end{center}
\end{figure}

At high sliding speed, $v=0.01$, the behavior is far more subtle, since now $\fric$ and $\Tsub$ both 
depend on $\gamma$. In addition, the time between stick-slip events is now much shorter: $\tauS = 
100$~LJ time units. For the anharmonic substrate, $\tauS$ still amply exceeds the lifetime $\tau_n$ 
of even the longest living phonon modes [\fig{fig5}]. We thus do not expect phonon backreflections 
to be important here. Instead, for the anharmonic substrate, the variation of friction with $\gamma$ 
should largely be due to the changing substrate temperature, $\Tsub$, via thermolubricity. In 
contrast, for the harmonic substrate, $\tauS$ is comparable to $\tau_n$, or even exceeds it 
[\fig{fig4}]. In this case, friction may well be affected by phonon backreflections, in addition to 
thermolubricity effects. To verify, we choose a fixed value of $\gamma$, then vary the thermostat 
temperature $\Tlan$, and monitor how $\fric$ changes with $\Tsub$ (this thus requires additional 
simulations to be performed, extending the range of $\Tlan$). For the harmonic substrate, for each 
value of $\gamma$, a different curve \ahum{$\fric$ versus $\Tsub$} is obtained [\fig{fig7}(a)]. In 
contrast, for the anharmonic substrate, the data for different $\gamma$ collapse onto a single curve 
[\fig{fig7}(b)].

\fig{fig7} strikingly confirms the expectations. For the anharmonic substrate, friction is set by 
the substrate temperature, $\Tsub$, the value of $\gamma$ being irrelevant here [\fig{fig7}(b)]. In 
addition, the decrease of $\fric$ with $\Tsub$ agrees with 
thermolubricity~\cite{10.1103/physrevlett.91.084502, 10.1103/physrevlett.104.256101}, the data being 
well described by $F_c - \fric \propto T^{2/3}$~\cite{10.1088/0953-8984/24/26/265001, 
10.1103/physrevlett.87.174301}, indicated by the dashed curve in \fig{fig7}(b). In contrast, for the 
harmonic substrate, \fig{fig7}(a) shows that $\fric$ depends on both $\Tsub$ and $\gamma$. If one 
increases $\Tsub$ keeping $\gamma$ fixed, $\fric$ decreases, similar to thermolubricity. Changing 
$\gamma$ affects both the strength of phonon backreflections, and the substrate temperature $\Tsub$, 
leading to a shift of the \ahum{$\fric$ versus $\Tsub$} curves.

Even though, for the harmonic substrate, backreflections are thus important, they are not strong 
enough to overturn thermolubricity. As \fig{fig2} shows, a strong anti-correlation between $\fric$ 
and $\Tsub$ remains clearly visible. In contrast, following \fig{fig4}, the correlation between 
$\fric$ and $\tau_n$ is much weaker (if backreflections were to dominate, $\fric$ should reach a 
minimum at $\gamma \sim 0.1-1$, since here $\tau_n$ is maximal). Hence, at least for the present 
data, the occurrence of the friction maximum with $\gamma$ appears to be mainly due to the reduction 
of $\Tsub$, rather than to the elimination of phonon backreflections.

\subsection{Role of thermostat}

We still mention one subtle point concerning the use of a (Langevin) thermostat. For the harmonic 
substrate, phonon lifetimes are extremely sensitive to $\gamma$ [cf.~\fig{fig4} and 
\fig{fig6}(a,c)]. These lifetimes thus mainly reflect the influence of the thermostat, i.e.~they are 
\ahum{biased}. An \ahum{unbiased} estimate could be obtained by removing the thermostat, which 
obviously can only be done in equilibrium, in which case the lifetime becomes maximally large 
[\fig{fig6}(a,c): star symbols]. Alas, under driving, removing the thermostat is not possible, and 
here one unavoidably simulates in a regime where the phonon dynamics of the entire substrate is 
largely dictated by the thermostat, which is against the notion of the thermostat being just a local 
perturbation, affecting only the lowest few substrate layers.

In contrast, for the anharmonic substrate, the phonon lifetime does not strongly depend on $\gamma$ 
[cf.~\fig{fig5} and \fig{fig6}(b,d)]. In this case, the lifetime is determined by the substrate 
particle interactions, thus reflecting a \ahum{true} (unbiased) material property. One could say 
that the anharmonic substrate is \ahum{self-thermalizing}: Due to the 3rd and 4th order terms in the 
particle interaction, phonon scattering is strong, and so the energy injected by the sliding bead at 
the top of the substrate, is quickly converted into heat. In order to control the temperature, all 
the thermostat has to do is remove this excess heat, but there is no need for the thermostat to 
actively generate heat, i.e.~a stochastic term is not needed. In fact, for the anharmonic substrate, 
even a simple velocity-rescaling thermostat~\cite{10.1063/1.2408420} suffices, which simply rescales 
the velocities of the particles being thermostatted every so often to match some chosen thermostat 
temperature $\Tlan$. In this case, varying now the frequency at which rescaling is performed, one 
still finds that $\Tsub$ deviates from $\Tlan$, but the curve of $\fric$ versus $\Tsub$ coincides 
with the one obtained using the Langevin thermostat [\fig{fig7}(b): hexagon symbols].

\section{Conclusions}

In this work, we have studied the problem of how to thermostat a nanotribology MD simulation, 
following-up on previous works~\cite{10.1103/physrevb.82.081401, 10.1007/s11249-012-9936-5, 
10.3762/bjnano.8.218}. The main insight of this work has been to relate thermostat effects to the 
lifetime of phonons in the underlying substrate. One finding is that, provided the sliding speed is 
sufficiently low, thermostat effects become negligible, the defining criterion being the typical 
phonon lifetime, which must be small compared to the time between stick-slip events. This result is 
convenient for modeling (clean) single-asperity AFM contacts, since here the use of a lower sliding 
speed would also better resemble experimental conditions~\cite{10.1103/physrevlett.106.126101}.

The opposite regime, where the phonon lifetime is comparable to or larger than the time between 
stick-slip events, could occur with single molecule adsorbates, due to the now much larger possible 
sliding velocity~\cite{10.1103/physrevb.59.16042}. In this regime, we confirm previous 
findings~\cite{10.1103/physrevb.82.081401, 10.1007/s11249-012-9936-5, 10.3762/bjnano.8.218}, namely, 
that the variation of friction with the Langevin damping parameter, $\gamma$, reveals a maximum. In 
addition, we also find that the substrate temperature, $\Tsub$, depends strongly on $\gamma$ then, 
reaching a minimum at the friction maximum. In this situation, we recommend plotting $\fric$ versus 
$\Tsub$. In case the latter yields a single curve, the $\gamma$ dependence could be a manifestation 
of thermolubricity~\cite{10.1103/physrevlett.91.084502, 10.1103/physrevlett.104.256101}, which one 
can verify by comparing to theoretical predictions~\cite{10.1088/0953-8984/24/26/265001, 
10.1103/physrevlett.87.174301}. The data may then be deemed to reliably describe friction, but at 
the substrate temperature, $\Tsub$, which here deviates from $\Tlan$ of the thermostat, typically 
exceeding it.

In contrast, when measurements of $\fric$ versus $\Tsub$ do not yield a single curve, one likely has 
long-lived backreflecting phonons in the system, which will strongly reduce friction. In this case, 
one could try to minimize them, by tuning $\gamma$ to the friction maximum, which follows the 
recommendation of earlier works~\cite{10.1103/physrevb.82.081401, 10.1007/s11249-012-9936-5, 
10.3762/bjnano.8.218}. The tuning of $\gamma$ has two effects, namely, (1) a reduction of the 
lifetime of backreflecting phonons, as shown by the vanishing of the vertical LA bands in 
\fig{fig4}, making these phonons more dissipating, and (2) a reduction of the substrate temperature, 
$\Tsub$, which increases friction due to thermolubricity. It is the combination of both effects that 
gives rise to the friction maximum.

Our phonon analysis also yields interesting insights into the underlying frictional mechanisms for a 
single sliding contact, at least for the present 2D model. Upon sliding, the energy of the bead 
appears to be mainly transferred into vertically propagating phonon modes (the \ahum{patterns} in 
\fig{fig4} and \fig{fig5} have a clear tendency to align in the $z$-direction). Depending now on how 
long these phonons live, this energy gets dissipated quickly to other phonon modes (short lifetime 
$\to$ large friction), or not (long lifetime $\to$ lower friction). Hence, frictional control 
tactics could be aimed at tuning the lifetimes of vertically propagating phonons. In situations 
where the phonon lifetime is long (i.e.~our harmonic substrate) using backreflections in this way, 
could potentially reduce friction by over 20\% [\fig{fig7}(a)].

We conclude with a word about system size effects. The present study used substrates containing 
$N=1600$ particles. Increasing $N$ will introduce new phonon modes into the system, which, depending 
on how these couple to the sliding bead, will likely affect friction (in fact, finite size effects 
in friction have been reported~\cite{10.1103/physrevb.82.081401, 10.1209/0295-5075/87/66002}). 
Further investigation of the interplay between system size and friction, indeed, how this might be 
exploited as control tactic, could be a topic for future work.

\acknowledgments

We acknowledge support by the German research foundation (SFB-1073, TP A01).

\bibliography{STUFF,AUTODOI}

\appendix

\section{Computation of phonon properties}

Central in the analysis of phonon properties is the dynamical matrix, which, for the 
2D hexagonal lattice, is a $2 \times 2$ matrix~\cite{10.1088/0143-0807/25/6/004, 
jones1985theoretical}
\begin{equation}
 \tilde{D}(\vec{k}) = \frac{4\alpha_2}{m} \sum_{\nu=0,1,2} [1-\cos(a \vec{k} \cdot \hat{n}_\nu)]
 \, \hat{n}_\nu \otimes \hat{n}_\nu \quad,
\end{equation}
with unit vectors $\hat{n}_\nu = (\cos \theta_\nu,\sin \theta_\nu)$, $\theta_\nu = 2\pi\nu/3$, $m$ 
the substrate particle mass, and $\alpha_2$ the coefficient of the quadratic term in the bond 
energy. We choose the wavevectors on a grid inside the FBZ, $\vec{k}(n_1,n_2) = (n_1 \vec{b}_1 + n_2 
\vec{b}_2) / n_{\rm max} + {\rm mod} (l_s \vec{d}_i)$, integers $0 \leq n_i < n_{\rm max}$, $n_{\rm 
max} = \sqrt{N}$, with $\vec{b}_i$ the reciprocal lattice vectors (which follow trivially from the 
real space primitive vectors $\vec{a}_i$ given in the main text). The {\it modulo} operation 
corresponds to repeated translations of length $l_s=4\pi/\sqrt{3}a$ (which is the height of the FBZ 
hexagon) in the directions $\vec{d}_i = \pm (\cos\phi_i, \sin\phi_i)$, $\phi_i \in \{ -\pi/6, \pi/6, 
\pi/2 \}$, until $\vec{k}$ lies inside the FBZ [\fig{fig1}(b)]. For each wavevector $\vec{k}$ inside 
the FBZ, diagonalization of $\tilde{D}(\vec{k})$ yields two eigenvalues, $\lambda_p(\vec{k})$, with 
corresponding eigenvectors $\hat{e}_p(\vec{k})$, normalized to unit length, where $p$ denotes the 
polarization. The eigenvalue yields the mode frequency, $\omega_p^2(\vec{k}) = \lambda_p(\vec{k})$; 
upon inspecting the inner product, $\vec{k} \cdot \hat{e}_p(\vec{k})$, one finds that the mode with 
the lowest frequency is predominantly transversal, the other longitudinal, and so we set $p \in \rm 
TA,LA$ in our notation. During the MD simulations, we record, for each particle $i=1,\ldots,N$, the 
displacement $\vec{u}_i(t)$ from its (time-averaged) position, and its velocity $\vec{v}_i(t)$, as 
function of time $t$. For this analysis, $\sim 20000$ time measurements were taken, each one 
separated by 10000 (1000) MD timesteps for the harmonic (anharmonic) substrate. From these data, we 
compute, for each wavevector $\vec{k}$ and polarization $p$, the normal mode coordinates at time 
$t$, $Q_p(\vec{k},t) = (\sqrt{m}/N) \sum_{i=1}^N \vec{u}_i(t) \cdot \hat{e}_p(\vec{k}) \, e^{-\imath 
\vec{k} \cdot \vec{R}_i}$, with an analogous expression for $\dot{Q}_p(\vec{k},t)$, where one 
replaces $\vec{u}_i (t) \to \vec{v}_i (t)$. In these equations, $\vec{R}_i$ is the perfect hexagonal 
lattice position of particle $i$ at the start of the simulation (i.e.~does not depend on $t$). The 
normal mode coordinates are then converted to mode amplitudes:
\begin{equation}
\begin{split}
 A_p(\vec{k},t) &= \\ 
 e^{\imath \omega_p(\vec{k}) t } &\sqrt{ \frac{N}{2\omega_p(\vec{k})} }
 \left( \omega_p(\vec{k}) Q_p(\vec{k},t) + \imath \dot{Q}_p(\vec{k},t) \right).
\end{split}
\end{equation}
Note that the normal mode coordinates and amplitudes are generally complex numbers. The 
instantaneous phonon energy is given by $E_p(\vec{k},t) = \omega_p(\vec{k})
|A_p(\vec{k},t)|^2$, which can be time-averaged over the MD trajectory to obtain $E(\vec{k},p)$. The 
phonon lifetime $\tau(\vec{k},p)$ is obtained from the autocorrelation function $\chi_p(\vec{k},t)$ 
of the corresponding amplitude time series $|A_p(\vec{k},t)|$. We use normalization 
$\chi_p(\vec{k},0)=1$; the time-averaged value of $|A_p(\vec{k},t)|$ is subtracted from the time 
series; fast Fourier transforms are used to compute $\chi_p(\vec{k},t)$. Precise estimates of 
$\tau(\vec{k},p)$ are difficult to obtain, since the functions $\chi_p(\vec{k},t)$ can be quite 
noisy. To this end, we first averaged $\chi_p(\vec{k},t)$ over all wavevectors, $\bar{\chi}_p (t) = 
(\sum_{\vec{k}} w_{\vec{k}} \chi_p(\vec{k},t)) / (\sum_{\vec{k}} w_{\vec{k}})$, with \ahum{weights} 
$w_{\vec{k}} = 1/|\vec{k}|$. The function $\bar{\chi}_p (t)$ decays to zero with increasing $t$ 
rather smoothly. We take the corresponding rate of the decay as a measure of the {\it typical} 
phonon lifetime, $\avg{\tau_p}$, shown in \fig{fig6}, which we compute using the integral measure: 
$\avg{\tau_p} = \int \bar{\chi}_p (t) \, dt$. To obtain the lifetime $\tau(\vec{k},p)$ of individual 
modes, we use the same integral measure, but apply it to the {\it envelope function} of 
$\chi_p(\vec{k},t)$, which one obtains via a Hilbert transform. The resulting estimates are scaled 
by a factor $f$ afterward, such that their (weighted) average matches the typical lifetime: 
$\avg{\tau_p} = (f \sum_{\vec{k}} w_{\vec{k}} \tau(\vec{k},p)) / (\sum_{\vec{k}} w_{\vec{k}})$, with 
$w_{\vec{k}}$ as given above.

\end{document}